%% file: Neural_Nano-Optics.tex
\documentclass{nature}

\usepackage[british]{babel}
\usepackage[utf8]{inputenc}
\usepackage{babelbib}
\usepackage{url}
\usepackage{graphicx}
\usepackage{subfig}
\usepackage{calc}
\usepackage{floatflt}
\usepackage{amssymb, amsmath, amsthm}
\usepackage{tabularx}
\usepackage{multirow}
\usepackage{array}
\usepackage{xcolor}
\usepackage{bm}
\usepackage{stmaryrd}
\usepackage{stackrel}
\usepackage{algpseudocode}
\usepackage{algorithm}
\usepackage{rotating}
\usepackage{mwe,tikz}\usepackage[percent]{overpic}
\usepackage{float}
\usepackage{listings}

\usepackage{numprint}
\usepackage{memhfixc}
\usepackage{makecell}
\usepackage[hidelinks]{hyperref}
\usepackage{siunitx}
\usepackage{textcomp}
\usepackage{lineno}
\linenumbers

\usepackage{mathtools}
\newcommand{\underExpl}[2]{%
  \underset{\substack{\uparrow\\\mathrlap{\text{\hspace{-1.5em}#2}}}}{#1}}
\newcommand{\overExpl}[2]{%
  \overset{\substack{\mathrlap{\text{\hspace{-1.5em}#2}}\\\downarrow}}{#1}}
\newcommand{\sL}{\hspace{2pt}(\hspace{2pt}}
\newcommand{\sR}{\hspace{2pt})\hspace{2pt}}

\title{Neural Nano-Optics for High-quality Thin Lens Imaging}

\author{Ethan Tseng,$^{1*}$ Shane Colburn,$^{2*}$ James Whitehead,$^2$ Luocheng Huang,$^2$ Seung-Hwan Baek,$^1$ Arka Majumdar,$^{2,3}$ Felix Heide$^{1\dagger}$}

\nolinenumbers

\begin{document}

\maketitle

\input{sections/definitions}

\begin{affiliations}
 \item Princeton University, Department of Computer Science
 \item University of Washington, Department of Electrical and Computer Engineering
 \item University of Washington, Department of Physics
 \item [$^*$] These authors contributed equally to this work
 \item [$^\dagger$] Corresponding author. E-mail: fheide@princeton.edu
\end{affiliations}

\begin{abstract}
\input{sections/abstract}
\end{abstract}



\input{sections/introduction}

\input{sections/results}

\input{sections/discussion}



\bibliographystyle{naturemag}
\bibliography{reference}


\begin{addendum}
    \item [Code and Data Availability] The code and data used to generate the findings of this study will be made public on GitHub. Throughout the review process, code and data are attached to this submission in the zip file named ``Neural\_Nano-Optics\_Code.zip''.
	\item This research was supported by NSF-1825308, DARPA (Contract no. 140D0420C0060), the UW Reality Lab, Facebook, Google, Futurewei, and Amazon. A.M. is also supported by a Washington Research Foundation distinguished investigator award. Part of this work was conducted at the Washington Nanofabrication Facility / Molecular Analysis Facility, a National Nanotechnology Coordinated Infrastructure (NNCI) site at the University of Washington with partial support from the National Science Foundation via awards NNCI-2025489 and NNCI-1542101.
	\item [Author Contributions] E.T. and F.H. developed the novel feature space-deconvolution technique, integrated the metasurface model and deconvolution framework, performed the final design optimizations, and led the manuscript writing. S.C. developed the differentiable metasurface and sensor noise model, led the experiment, and assisted E.T. in writing the manuscript. J.W. fabricated all the devices and assisted in the experiment. L.H. developed the scripts for automated image capture and assisted in the experiment. S.B. assisted in writing the manuscript. A.M. and F.H. supervised the project and assisted in writing the manuscript.
	\item [Competing Interests] A.M. is co-founder of Tunoptix Inc., which is commercializing technology discussed in this manuscript.
	\item [Supplementary Information] Supplementary Information is attached to this submission.
	\item [Correspondence] Correspondence should be addressed to Felix Heide.
\end{addendum}

\input{sections/figures}


\end{document}

%% file: sections/definitions.tex

\definecolor{Gray}{rgb}{0.5,0.5,0.5}
\definecolor{darkblue}{rgb}{0,0,0.7}
\definecolor{orange}{rgb}{1,.5,0} 
\definecolor{red}{rgb}{1,0,0} 

\newcommand{\heading}[1]{\noindent\textbf{#1}}
\newcommand{\note}[1]{{{\textcolor{orange}{#1}}}}
\newcommand{\todo}[1]{{\textcolor{darkblue}{TODO: #1}}}
\newcommand{\changed}[1]{{\textcolor{blue}{#1}}}
\newcommand{\revision}[1]{{{#1}}}
\newcommand{\place}[1]{ \begin{itemize}\item\textcolor{darkblue}{#1}\end{itemize}}
\newcommand{\de}{\mathrm{d}}

\newcommand{\BEAS}{\begin{eqnarray*}}
\newcommand{\EEAS}{\end{eqnarray*}}
\newcommand{\BEA}{\begin{eqnarray}}
\newcommand{\EEA}{\end{eqnarray}}
\newcommand{\BEQ}{\begin{equation}}
\newcommand{\EEQ}{\end{equation}}
\newcommand{\BIT}{\begin{itemize}}
\newcommand{\EIT}{\end{itemize}}
\newcommand{\BNUM}{\begin{enumerate}}
\newcommand{\ENUM}{\end{enumerate}}

\newcommand{\BA}{\begin{array}}
\newcommand{\EA}{\end{array}}

\newcommand{\eg}{{\it e.g.}}
\newcommand{\ie}{{\it i.e.}}
\newcommand{\etc}{{\it etc.}}

\newcommand{\ones}{\mathbf 1}

\newcommand{\reals}{{\mbox{\bf R}}}
\newcommand{\integers}{{\mbox{\bf Z}}}
\newcommand{\eqbydef}{\mathrel{\stackrel{\Delta}{=}}}
\newcommand{\complex}{{\mbox{\bf C}}}
\newcommand{\symm}{{\mbox{\bf S}}}  

\newcommand{\Span}{\mbox{\textrm{span}}}
\newcommand{\Range}{\mbox{\textrm{range}}}
\newcommand{\nullspace}{{\mathcal N}}
\newcommand{\range}{{\mathcal R}}
\newcommand{\Nullspace}{\mbox{\textrm{nullspace}}}
\newcommand{\Rank}{\mathop{\bf Rank}}
\newcommand{\Tr}{\mathop{\bf Tr}}
\newcommand{\diag}{\mathop{\bf diag}}
\newcommand{\lambdamax}{{\lambda_{\rm max}}}
\newcommand{\lambdamin}{\lambda_{\rm min}}

\newcommand{\Expect}{\mathop{\bf E{}}}
\newcommand{\Prob}{\mathop{\bf Prob}}
\newcommand{\erf}{\mathop{\bf erf}}

\newcommand{\Co}{{\mathop {\bf Co}}}
\newcommand{\co}{{\mathop {\bf Co}}}
\newcommand{\dist}{\mathop{\bf dist{}}}
\newcommand{\Ltwo}{{\bf L}_2}
\newcommand{\QED}{~~\rule[-1pt]{8pt}{8pt}}\def\qed{\QED}
\newcommand{\approxleq}{\mathrel{\smash{\makebox[0pt][l]{\raisebox{-3.4pt}{\small$\sim$}}}{\raisebox{1.1pt}{$<$}}}}
\newcommand{\epi}{\mathop{\bf epi}}

\newcommand{\vol}{\mathop{\bf vol}}
\newcommand{\Vol}{\mathop{\bf vol}}
\newcommand{\Card}{\mathop{\bf card}}

\newcommand{\dom}{\mathop{\bf dom}}
\newcommand{\aff}{\mathop{\bf aff}}
\newcommand{\cl}{\mathop{\bf cl}}
\newcommand{\Angle}{\mathop{\bf angle}}
\newcommand{\intr}{\mathop{\bf int}}
\newcommand{\relint}{\mathop{\bf rel int}}
\newcommand{\bd}{\mathop{\bf bd}}
\newcommand{\vect}{\mathop{\bf vec}}
\newcommand{\dsp}{\displaystyle}
\newcommand{\foequal}{\simeq}
\newcommand{\VOL}{{\mbox{\bf vol}}}
\newcommand{\xopt}{x^{\rm opt}}

\newcommand{\Xb}{{\mbox{\bf X}}}
\newcommand{\xst}{x^\star}
\newcommand{\varphist}{\varphi^\star}
\newcommand{\lambdast}{\lambda^\star}
\newcommand{\Zst}{Z^\star}
\newcommand{\fstar}{f^\star}
\newcommand{\xstar}{x^\star}
\newcommand{\xc}{x^\star}
\newcommand{\lambdac}{\lambda^\star}
\newcommand{\lambdaopt}{\lambda^{\rm opt}}

\newcommand{\geqK}{\mathrel{\succeq_K}}
\newcommand{\gK}{\mathrel{\succ_K}}
\newcommand{\leqK}{\mathrel{\preceq_K}}
\newcommand{\lK}{\mathrel{\prec_K}}
\newcommand{\geqKst}{\mathrel{\succeq_{K^*}}}
\newcommand{\gKst}{\mathrel{\succ_{K^*}}}
\newcommand{\leqKst}{\mathrel{\preceq_{K^*}}}
\newcommand{\lKst}{\mathrel{\prec_{K^*}}}
\newcommand{\geqL}{\mathrel{\succeq_L}}
\newcommand{\gL}{\mathrel{\succ_L}}
\newcommand{\leqL}{\mathrel{\preceq_L}}
\newcommand{\lL}{\mathrel{\prec_L}}
\newcommand{\geqLst}{\mathrel{\succeq_{L^*}}}
\newcommand{\gLst}{\mathrel{\succ_{L^*}}}
\newcommand{\leqLst}{\mathrel{\preceq_{L^*}}}
\newcommand{\lLst}{\mathrel{\prec_{L^*}}}

\newtheorem{theorem}{Theorem}[section]
\newtheorem{corollary}{Corollary}[theorem]
\newtheorem{lemma}[theorem]{Lemma}
\newtheorem{proposition}[theorem]{Proposition}

\newenvironment{algdesc}%
{\begin{quote}}{\end{quote}}

\def\figbox#1{\framebox[\hsize]{\hfil\parbox{0.9\hsize}{#1}}}

\makeatletter
\long\def\@makecaption#1#2{
   \vskip 9pt
   \begin{small}
   \setbox\@tempboxa\hbox{{\bf #1:} #2}
   \ifdim \wd\@tempboxa > 5.5in
        \begin{center}
        \begin{minipage}[t]{5.5in}
        \addtolength{\baselineskip}{-0.95pt}
        {\bf #1:} #2 \par
        \addtolength{\baselineskip}{0.95pt}
        \end{minipage}
        \end{center}
   \else
    \hbox to\hsize{\hfil\box\@tempboxa\hfil}
   \fi
   \end{small}\par
}
\makeatother

\newcounter{oursection}
\newcommand{\oursection}[1]{
 \addtocounter{oursection}{1}
 \setcounter{equation}{0}
 \clearpage \begin{center} {\Huge\bfseries #1} \end{center}
 {\vspace*{0.15cm} \hrule height.3mm} \bigskip
 \addcontentsline{toc}{section}{#1}
}
\newcommand{\oursectionf}[1]{  
 \addtocounter{oursection}{1}
 \setcounter{equation}{0}
 \foilhead[-.5cm]{#1 \vspace*{0.8cm} \hrule height.3mm }
 \LogoOn
}
\newcommand{\oursectionfl}[1]{  
 \addtocounter{oursection}{1}
 \setcounter{equation}{0}
 \foilhead[-1.0cm]{#1}
 \LogoOn
}

\newcommand{\Mat}[1]    {{\ensuremath{\mathbf{\uppercase{#1}}}}} 
\newcommand{\Vect}[1]   {{\ensuremath{\mathbf{\lowercase{#1}}}}} 
\newcommand{\Vari}[1]   {{\ensuremath{\mathbf{\lowercase{#1}}}}} 
\newcommand{\Id}				{\mathbb{I}} 
\newcommand{\Diag}[1] 	{\operatorname{diag}\left({ #1 }\right)} 
\newcommand{\Opt}[1] 	  {{#1}_{\text{opt}}} 
\newcommand{\CC}[1]			{{#1}^{*}} 
\newcommand{\Op}[1]     {\Mat{#1}} 
\newcommand{\mini}[1] {{\mbox{argmin}}_{#1} \: \: } 
\newcommand{\argmin}[1] {\underset{{#1}}{\mathop{\rm argmin}} \: \: } 
\newcommand{\argmax}[1] {\underset{{#1}}{\mathop{\rm argmax}} \: \: } 
\newcommand{\minimize}{\mathop{\rm minimize} \: \:}
\newcommand{\minimizeu}[1]{\underset{{#1}}{\mathop{\rm minimize}} \: }
\newcommand{\grad}      {\nabla}
\newcommand{\kron}{\otimes} 

\newcommand{\gradt}     {\grad_\z}
\newcommand{\gradx}     {\grad_\x}
\newcommand{\Drv}     	{\Mat{D}} 
\newcommand{\step}      {\text{\textbf{step}}}
\newcommand{\prox}[1]   {\mathbf{prox}_{#1}}
\newcommand{\ind}[1]    {\operatorname{ind}_{#1}}
\newcommand{\proj}[1]   {\Pi_{#1}}
\newcommand{\pointmult}{\odot} 
\newcommand{\rr}   {\mathcal{R}}

\newcommand{\Basis}{\Mat{D}}         		
\newcommand{\Corr}{\Mat{C}}             
\newcommand{\conv}{\ast} 
\newcommand{\meas}{\Vect{b}}            
\newcommand{\Img}{I}                    
\newcommand{\img}{\Vect{i}}             
\newcommand{\vv}{\Vect{v}}
\newcommand{\p}{\Vect{p}}
\newcommand{\Splitvar}{T}                
\newcommand{\splitvar}{\Vect{t}}         
\newcommand{\Splitbasis}{J}                
\newcommand{\splitbasis}{\Vect{j}}         
\newcommand{\var}{\Vari{z}} 

\newcommand{\FT}[1]			{\mathcal{F}\left( {#1} \right)} 
\newcommand{\IFT}[1]			{\mathcal{F}^{-1}\left( {#1} \right)} 

\newcommand{\func}{f}
\newcommand{\fMat}{\Mat{K}}

\newcommand{\avar}{\Vari{v}} 
\newcommand{\aspvar}{\Vari{z}} 

\newcommand{\mask}{\Mat{M}}

\newcommand{\Pen}      		{F} 
\newcommand{\cardset}     {\mathcal{C}}
\newcommand{\Dat}      		{G} 
\newcommand{\Reg}      		{\Gamma} 

\newcommand{\Trans}{\mathbf{\uppercase{T}}} 
\newcommand{\Ph}{\mathbf{\uppercase{\Phi}}} 

\newcommand{\Tvec}{\Vect{T}} 
\newcommand{\Bvec}{\Vect{B}} 

\newcommand{\Wt}{\Mat{W}} 

\newcommand{\Perm}{\Mat{P}} 
\newcommand{\C}{\Mat{C}} 

\newcommand{\DiagFactor}[1]     {\Mat{O}_{ #1 }}  

\newcommand{\Proj}{\Mat{P}}             

\newcommand{\Vector}[1]{\mathbf{#1}}
\newcommand{\Matrix}[1]{\mathbf{#1}}
\newcommand{\Tensor}[1]{\boldsymbol{\mathscr{#1}}}
\newcommand{\TensorUF}[2]{\Matrix{#1}_{(#2)}}

\newcommand{\MatrixKP}[1]{\Matrix{#1}_{\otimes}}
\newcommand{\MatrixKPN}[2]{\Matrix{#1}_{\otimes}^{#2}}

\newcommand{\MatrixKRP}[1]{\Matrix{#1}_{\odot}}
\newcommand{\MatrixKRPN}[2]{\Matrix{#1}_{\odot}^{#2}}

\newcommand{\HP}{\circ}
\newcommand{\HD}{\oslash}

\newcommand{\leftDB}{\left[ \! \left[}
\newcommand{\rightDB}{\right] \! \right]}

\newcommand{\transpose}{T}

\newcommand*\sstrut[1]{\vrule width0pt height0pt depth#1\relax}

\newcommand{\inlineeqnum}{\refstepcounter{equation}~~\mbox{(\theequation)}}
\newcommand{\eqname}[1]{\tag*{#1~(\theequation)}\refstepcounter{equation}}

\newcommand{\lambdas}{\boldsymbol{\lambda}}
\newcommand{\alb}{\boldsymbol{\alpha}} 	
\newcommand{\depth}{\boldsymbol{z}} 	
\newcommand{\albi}{\alpha} 	
\newcommand{\depthi}{z} 	
\newcommand{\ambient}{s}
\newcommand{\jitter}{w}
\newcommand{\z}{\Vect{z}} 							
\newcommand{\x}{\Vect{x}}             	
\newcommand{\y}{\Vect{y}}             	
\newcommand{\Kvar}{\Mat{K}}
\newcommand{\lagrangemult}{\boldsymbol{\nu}}
\newcommand{\scaledlagrange}{\Vect{u}}
\newcommand{\eps}{\epsilon}
\newcommand{\vp}{\Vect{v}}

%% file: sections/abstract.tex
Nano-optic imagers that modulate light at sub-wavelength scales could unlock unprecedented applications in diverse domains ranging from robotics to medicine. Although metasurface optics offer a path to such ultra-small imagers, existing methods have achieved image quality far worse than bulky refractive alternatives, fundamentally limited by aberrations at large apertures and low f-numbers. In this work, we close this performance gap by presenting the first \textit{neural nano-optics}. We devise a fully differentiable learning method that learns a metasurface physical structure in conjunction with a novel, neural feature-based image reconstruction algorithm. Experimentally validating the proposed method, we achieve an order of magnitude lower reconstruction error. As such, we present the first high-quality, nano-optic imager that combines the widest field of view for full-color metasurface operation while simultaneously achieving the largest demonstrated 0.5 mm, f/2 aperture.

%% file: sections/introduction.tex
The miniaturization of intensity sensors in recent decades has made today's cameras ubiquitous across many application domains, including medical imaging, commodity smartphones, security, robotics, and autonomous driving. However, only imagers that are an order of magnitude smaller could enable novel applications in nano-robotics, {\it in vivo} imaging, AR/VR, and health monitoring. While sensors with sub-micron pixels do exist, further miniaturization has been prohibited by fundamental limitations of conventional optics. Traditional systems consist of a cascade of refractive elements that correct for aberrations, and these bulky lenses impose a lower limit on camera footprint. A further fundamental barrier is the difficulty of reducing focal length, as this induces greater chromatic aberrations.

We turn towards computationally designed metasurface optics (meta-optics) to close this gap and enable ultra-compact cameras that could allow for unprecedented capabilities in endoscopy, brain imaging, or in a distributed fashion as collaborative optical ``dust'' on scene surfaces. Ultrathin meta-optics utilize subwavelength nano-antennas to modulate incident light with greater design freedom and space-bandwidth product over conventional diffractive optical elements (DOEs)\cite{Engelberg2020TheAO,Lin298Dielectric,Wetzstein2020InferenceIA,Peng2019LearnedLF}. Researchers have harnessed this potential for building flat optics for imaging\cite{Yu2014FlatOW,Aieta2012AberrationFree}, polarization control\cite{Arbabi2015DielectricMF}, and holography\cite{Zheng2015MetasurfaceHR}. Existing metasurface imaging methods, however, achieve an order of magnitude higher reconstruction error than achievable with refractive compound lenses due to severe, wavelength-dependent aberrations that arise from discontinuities in their imparted phase\cite{Colburn2018MetasurfaceOF,Yu2014FlatOW,Lin298Dielectric,Arbabi2016MiniatureOP,Avayu2017CompositeFM,Aieta2015MultiwavelengthAM,Khorasaninejad2016MetalensesAV,Wang2018ABA,Shrestha2018BroadbandAD}. Dispersion-engineered metasurfaces aims to mitigate this by exploiting group delay and group delay dispersion to focus broadband light\cite{Ndao2020OctaveBP,Chen2018ABA,Shrestha2018BroadbandAD,Wang2018ABA,Arbabi2017ControllingTS,khorasaninejad2017achromatic,wang2017broadband}, but this technique is fundamentally limited\cite{Presutti2020FocusingLimits}, constraining designs to apertures of ${\approx}{10}$'s of microns. As such, existing approaches have not been able to increase the achievable aperture sizes without significantly reducing the numerical aperture or supported wavelength range. Other attempted solutions only suffice for discrete wavelengths or narrowband illumination\cite{Aieta2015MultiwavelengthAM,Avayu2017CompositeFM,Wang2016VisibleFrequencyDM,Arbabi2016MiniatureOP,Khorasaninejad2016MetalensesAV}.

Metasurfaces also exhibit strong geometric aberrations that have limited their utility for wide field of view (FOV) imaging. Approaches that support wide FOV typically rely on either small input apertures that limit light collection\cite{Shalaginov2019ASP} or use multiple metasurfaces\cite{Arbabi2016MiniatureOP}, which drastically increases fabrication complexity. Moreover, these multiple metasurfaces are separated by a gap that scales linearly with the aperture, thus obviating the size benefit of meta-optics as the aperture increases.

Recently, researchers have leveraged computational imaging to offload aberration correction to post-processing software\cite{Colburn2018MetasurfaceOF,Colburn2020SimultaneousAA,Heide2013HighqualityCI}.
Although these approaches enable full-color imaging metasurfaces without stringent aperture limitations, they are limited to a FOV below ${20^\circ}$ and the reconstructed spatial resolution is an order of magnitude below that of conventional refractive optics.

Researchers have similarly proposed camera designs that utilize a single optic instead of compound stacks\cite{Peng2015ComputationalIU,Peng2016DiffractiveAchromat}, but these systems fail to match the performance of commodity imagers due to low diffraction efficiency. Moreover, the most successful approaches\cite{heide2016encoded,Peng2015ComputationalIU,Peng2016DiffractiveAchromat} hinder miniaturization because of their long backfocal distances of more than $\SI{10}{mm}$. Lensless cameras\cite{White2020ASP} instead reduce size by replacing the optics with amplitude masks, but this severely limits spatial resolution and requires long acquisition times.

In this work, we propose \textit{neural nano-optics}, leveraging a learned design method that overcomes these limitations of existing techniques. In contrast to previous works that rely on hand-crafted designs, we co-optimize the metasurface and deconvolution algorithm with an end-to-end differentiable model of image formation and computational reconstruction. This model exploits a memory-efficient differentiable nano-scatterer model, as well as a novel, neural feature-based reconstruction architecture. The jointly optimized nano-optic can be fabricated at a mass scale using deep ultraviolet (DUV) lithography. Our approach departs from inverse-designed meta-optics\cite{mansouree2020multifunctional,Chung2020HighNAAM} in that we support larger aperture sizes and directly optimize the quality of the final image as opposed to intermediate metrics such as the focal spot intensity.
Although end-to-end optimization of DOEs has been explored\cite{stork2014optical,Sitzmann2018EndtoendOO,Sun2020LearnedOpticHDR}, existing methods using phase plates assume shift-invariant systems and only support small FOVs of ${\approx} {5^\circ}$. Furthermore, existing learned deconvolution methods are only minor variations of standard encoder-decoder architectures, such as the U-Net\cite{Ronneberger2015UNetCN}, and often fail to generalize to experimental measurements or handle large spatially-dependent aberrations, as found in metasurface images.

With the proposed neural nano-optics, we achieve the first high-quality, polarization-insensitive nano-optic imager for full-color ($\SI{400}{nm}$ to $\SI{700}{nm}$), wide FOV ($40^\circ$) imaging with an f-number of $2$. For our $\SI{500}{\micro m}$ aperture, we optimized $1.6 \times 10^6$ nano-scatterers, which is an order of magnitude greater than existing achromatic meta-optics. Compared to all existing heuristically designed metasurfaces and metasurface computational imaging approaches, we outperform existing methods by an order of magnitude in reconstruction error outside the nominal wavelength range on experimental captures.

%% file: sections/results.tex
\subsection{Differentiable Metasurface Proxy Model}
The proposed differentiable metasurface image formation model (Fig.~\ref{fig:pipeline}) consists of three sequential stages that utilize differentiable tensor operations: metasurface phase determination, PSF simulation and convolution, and sensor noise. In our model, polynomial coefficients that determine the metasurface phase are optimizable variables, whereas experimentally calibrated parameters characterizing the sensor readout and the sensor-metasurface distance are fixed.

The optimizable metasurface phase function $\phi$ as a function of distance $r$ from the optical axis is given by
\begin{equation}
\phi(r) = \sum_{i=0}^n a_i \left(\frac{r}{R}\right)^{2i},
\end{equation}
where the $\{a_0, \ldots a_n\}$ are optimizable coefficients, $R$ is the phase mask radius, and $n$ is the number of polynomial terms. We optimize the metasurface in this phase function basis as opposed to in a pixel-by-pixel manner to avoid local minima, see Supplemental Material.
This phase, however, is only defined for a single, nominal design wavelength, which is a fixed hyperparameter in our optimization. While this mask alone is sufficient for modeling monochromatic light propagation, we require the phase at all target wavelengths to design for a broadband imaging scenario.

To this end, at each position in our metasurface we apply two sequential operations. The first operation is an inverse, phase-to-structure mapping that computes the scatterer geometry given the desired phase at the nominal design wavelength. With the scatterer geometry determined, we can then apply a forward, structure-to-phase mapping to calculate the phase at the remaining target wavelengths. Leveraging an effective index approximation that ensures a unique geometry for each phase shift in the 0 to $2\pi$ range, we ensure differentiability, and can directly optimize the phase coefficients by adjusting the scatterer dimensions and computing the response at different target wavelengths, see Supplemental Document.

These phase distributions differentiably determined from the nano-scatterers allow us to then calculate the PSF as a function of wavelength and field angle to efficiently model full-color image formation over the whole FOV, see Supplemental Document. Finally, we simulate sensing and readout with experimentally calibrated Gaussian and Poisson noise by using the reparameterization and score-gradient techniques to enable backpropagation, see Supplemental Document.

When compared directly against alternative computational forward simulation methods, such as finite-difference time-domain (FDTD) simulation\cite{mansouree2020multifunctional}, our technique is approximate but is more than three orders of magnitudes faster and more memory efficient. For the same aperture as our design, FDTD simulation would require on the order of 30 terabytes for accurate meshing alone. Our technique instead only scales quadratically with length. This enables our entire end-to-end pipeline to achieve a \emph{memory reduction of over $3000 \times$}, with metasurface simulation and image reconstruction both fitting within a few gigabytes of GPU RAM.

\subsection{Neural Feature Propagation and Learned Nano-Optics Design}
We propose a neural deconvolution method that incorporates learned priors while generalizing to unseen test data. Specifically, we design a neural network architecture that performs deconvolution on a learned feature space instead of on raw image intensity. This technique combines both the generalization of model-based deconvolution and the effective feature learning of neural networks, allowing us to tackle image deconvolution for meta-optics with severe aberrations and spatially large PSFs. This approach generalizes well to experimental captures even when trained only in simulation, see Supplemental Document.

The proposed reconstruction network architecture comprises three stages: a multi-scale feature extractor $f_\textsc{fe}$, a propagation stage $f_{\textsc{z} \rightarrow \textsc{w}}$ that deconvolves these features (i.e., propagates features $\textsc{Z}$ to their deconvolved spatial positions $\textsc{W}$), and a decoder stage $f_\textsc{de}$ that combines the propagated features into a final image. Formally, our feature propagation network performs the following operations:
\begin{equation}
\mathbf{O} = \underExpl{f_\textsc{de}}{Decoder}\sL\overExpl{f_{\textsc{z} \rightarrow \textsc{w}}}{Feature Propagation}\sL\underExpl{f_\textsc{fe}}{Feature Extraction}\sL\mathbf{I}\sR, \hspace{1pt}\text{PSF}\sR),
\end{equation}
where $\mathbf{I}$ is the raw sensor measurement and $\mathbf{O}$ is the output image.

Both the feature extractor and decoder are constructed as fully convolutional neural networks. The feature extractor identifies features at both the native resolution and multiple scales to facilitate learning low-level and high-level features, allowing us to encode and propagate higher-level information beyond raw intensity. The subsequent feature propagation stage $f_{\textsc{z} \rightarrow \textsc{w}}$ then propagates the features to their inverse-filtered positions using a differentiable deconvolution method. Finally, the decoder stage then converts the propagated features back into image space, see Supplemental Document for architecture details. When compared against existing state-of-the-art deconvolution approaches we achieve over $\SI{4}{dB}$ PSNR improvement (more than $2.5\times$ reduction in mean squared error) for deconvolving challenging metasurface incurred aberrations, see Supplemental Document.

Both our metasurface image formation model and our deconvolution algorithm are incorporated into a fully differentiable, end-to-end imaging chain. Our metasurface imaging pipeline allows us to apply first-order stochastic optimization methods to learn metasurface phase parameters $\mathcal{P}_\textsc{meta}$ and parameters $\mathcal{P}_\textsc{deconv}$ for our deconvolution network $f_\textsc{deconv}$ that will minimize our endpoint loss function $\mathcal{L}$, which in our case is a perceptual quality metric. Our image formation model is thus defined as 
\begin{equation}
\mathbf{O} = f_\textsc{deconv}(\mathcal{P}_\textsc{deconv}, f_\textsc{sensor}(\mathbf{I} * f_\textsc{meta}(\mathcal{P}_\textsc{meta})), f_\textsc{meta}(\mathcal{P}_\textsc{meta}))),
\end{equation}
where $\mathbf{I}$ is an RGB training image, $f_\textsc{meta}$ generates the metasurface PSF from $\mathcal{P}_\textsc{meta}$, $*$ is convolution, and $f_\textsc{sensor}$ models the sensing process including sensor noise. Since our deconvolution method is non-blind, $f_\textsc{deconv}$ takes in $f_\textsc{meta}(\mathcal{P}_\textsc{meta})$. We then solve the following optimization problem
\begin{equation}
\{\mathcal{P}^*_\textsc{meta}, \mathcal{P}^*_\textsc{deconv}\} = \argmin{\mathcal{P}_\textsc{meta}, \mathcal{P}_\textsc{deconv}} \sum_{i=1}^M \mathcal{L} (\mathbf{O}^{(i)} , \mathbf{I}^{(i)}).
\end{equation}
The final learned parameters $\mathcal{P}^*_\textsc{meta}$ are used to manufacture the meta-optic and $\mathcal{P}^*_\textsc{deconv}$ determines the deconvolution algorithm, see Supplemental Material for further details.

\subsection{Imaging Demonstration}
High-quality, full-color image reconstructions using our neural nano-optic are shown in Fig.~\ref{fig:results} and the Supplemental Document. We perform comparisons against a traditional hyperbolic meta-optic designed for $\SI{511}{nm}$ and the state-of-the-art cubic meta-optic from Colburn et al.\cite{Colburn2018MetasurfaceOF}. Additional comparisons against alternative single-optic and meta-optic designs are shown in the Supplemental Document. Ground truth images are acquired using a six-element compound optic that is \emph{550000{$\times$} larger in volume than the meta-optics.} Our full computational reconstruciton pipeline runs at real-time rates and requires only $\SI{58}{ms}$ to process a $\SI{720}{px} \times \SI{720}{px}$ RGB capture.

The traditional hyperbolic meta-optic experiences severe chromatic aberrations at larger and shorter wavelengths. This is observed in the heavy red blurring in Fig.~\ref{fig:results}(A) and the washed out blue color in Fig.~\ref{fig:results}(C). The cubic meta-optic maintains better consistency across color channels but suffers from artifacts owing to its large, asymmetric PSF. In contrast, we demonstrate high-quality images without these aberrations, which are observable in the fine details in the fruits in Fig.~\ref{fig:results}(A), the patterns on the lizard in Fig.~\ref{fig:results}(B), and the flower petals in Fig.~\ref{fig:results}(C). We quantitatively validate the proposed neural nano-optic by measuring reconstruction error on an unseen test set of natural images, on which we obtain $10{\times}$ lower mean-squared error than existing approaches, see Supplemental Document. In addition to natural image reconstruction, we also measured the spatial resolution using standard test charts, see Supplemental Document. Our nano-optic imager achieves an image-side spatial resolution of $\SI{214}{lp/mm}$ across all color channels at $\SI{120}{mm}$ object distance. We improve spatial resolution by an order of magnitude over the previous state-of-the-art by Colburn et al.\cite{Colburn2018MetasurfaceOF} which achieved $\SI{30}{lp/mm}$.

\subsection{Characterizing Nano-Optics Performance}
Through our optimization process, our meta-optic learns to produces compact spatial PSFs that minimize chromatic aberrations across all color channels. Unlike designs that exhibit a sharp focus for a single wavelength but significant aberrations at other wavelengths, our optimized design strikes a balance across wavelengths to facilitate full-color imaging. Furthermore, the learned metasurface avoids the spatially large PSFs used previously by Colburn et al.\cite{Colburn2018MetasurfaceOF} for computational imaging.

After optimization, we fabricated our neural meta-optics (Fig.~\ref{fig:fab}), as well as several heuristic designs for comparison, see Supplemental Document. Note that commercial large-scale production of our nano-optic can be performed using high-throughput processes based on DUV lithography which is standard for mature industries such as integrated circuits. The simulated and experimental PSFs are shown in Fig.~\ref{fig:fab} and are in strong agreement, validating the physical accuracy of the proxy metasurface model. To account for manufacturing imperfections, we perform a PSF calibration step where we capture the spatial PSFs using the fabricated optics. We then finetune our deconvolution network by replacing the proxy-based metasurface simulator with the captured PSFs. The finetuned network is deployed on experimental captures using the setup shown in Fig. S8. This finetuning calibration step does not train on experimental captures, we only require the measured PSFs, without requiring experimental collection of a vast image dataset.

We observe that the PSF for our optimized meta-optic exhibits a combination of compact shape and minimal variance across field angles, as expected for our design. PSFs for a traditional hyperbolic meta-optic ($\SI{511}{nm}$) instead have significant spatial variation across field angles and severe chromatic aberrations that cannot be compensated through deconvolution. While the cubic design from Colburn et al.\cite{Colburn2018MetasurfaceOF} does exhibit spatial invariance, its asymmetry and large spatial extent introduce severe artifacts that reduce image quality. See Fig.~\ref{fig:fab} and Supplemental Document for comparisons of the traditional meta-optic and Colburn et al.\cite{Colburn2018MetasurfaceOF} against ours. We also show corresponding modulation transfer functions (MTFs) for our design in Fig.~\ref{fig:fab}. The MTF does not change appreciably with incidence angle and also preserves a broad range of spatial frequencies across the visible spectrum.

%% file: sections/discussion.tex
\subsection{Discussion}
In this work, we present a new paradigm for achieving high-quality, full-color, wide FOV imaging using \textit{neural nano-optics}. Specifically, the proposed learned imaging method allows for an order of magnitude lower reconstruction error on experimental data than existing works.
The key enablers of this result are our differentiable meta-optical image formation model and novel deconvolution algorithm.
Combined together as a differentiable end-to-end model, we jointly optimize the full computational imaging pipeline with the only target metric being the quality of the deconvolved RGB image -- sharply deviating from existing methods that penalize focal spot size in isolation from the reconstruction method.

We have demonstrated, for the first time, the viability of meta-optics for high-quality imaging in full-color, over a wide FOV. No existing meta-optic demonstrated to date approaches a comparable combination of image quality, large aperture size, low f-number, wide fractional bandwidth, wide FOV, and polarization insensitivity (see Supplemental Document), and the proposed method could scale to mass production. Furthermore, we demonstrate \emph{image quality on par with a bulky, six-element commercial compound lens even though our design volume is 550000{$\times$} lower and utilizes a single metasurface}.

We have designed \textit{neural nano-optics} for a dedicated imaging task, but we envision extending our work towards flexible imaging with reconfigurable nanophotonics for diverse tasks, ranging from extended depth of field to classification or object detection tasks. We believe that the proposed method takes an essential step towards ultra-small cameras that may enable novel applications in endoscopy, brain imaging, or in a distributed fashion on scene surfaces.

%% file: sections/figures.tex
\begin{figure*}
    \centering
    \includegraphics[width=0.99\linewidth]{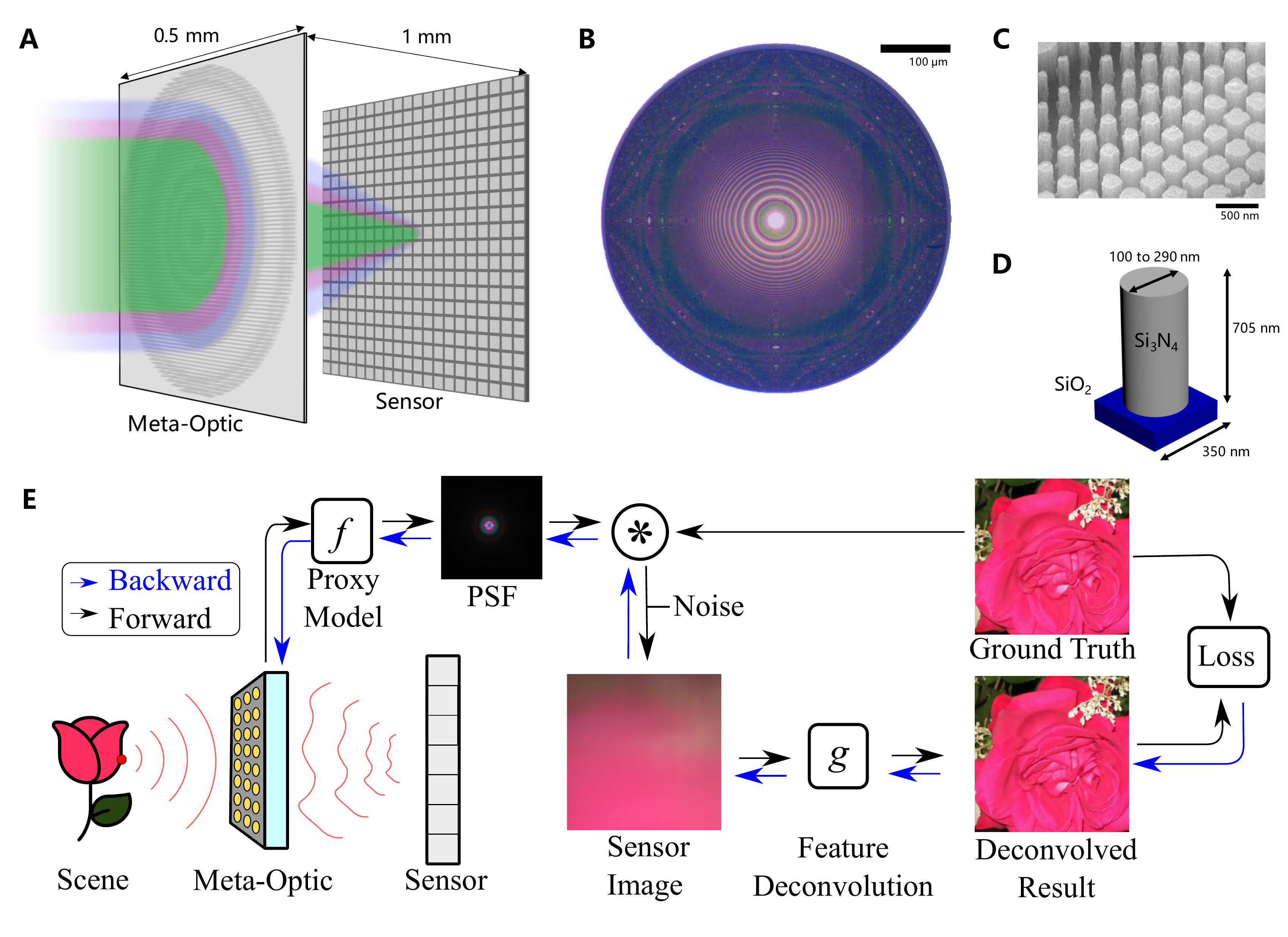}
    \caption{Our learned, ultrathin meta-optic as shown in (A) is $\SI{500}{\micro m}$ in thickness and diameter, allowing for the design of a miniature camera. The manufactured optic is shown in (B). A zoom-in is shown in (C) and nanopost dimensions are shown in (D). Our end-to-end imaging pipeline shown in (E) is composed of the proposed efficient metasurface image formation model and the feature-based deconvolution algorithm. From the optimizable phase profile our differentiable model produces spatially-variant PSFs, which are then {patch-wise} convolved with the input image to form the sensor measurement. The sensor reading is then deconvolved using our algorithm to produce the final image.}
    \label{fig:pipeline}
\end{figure*}

\begin{figure*}
    \vspace{-24pt}
    \centering
    \includegraphics[width=0.9\linewidth]{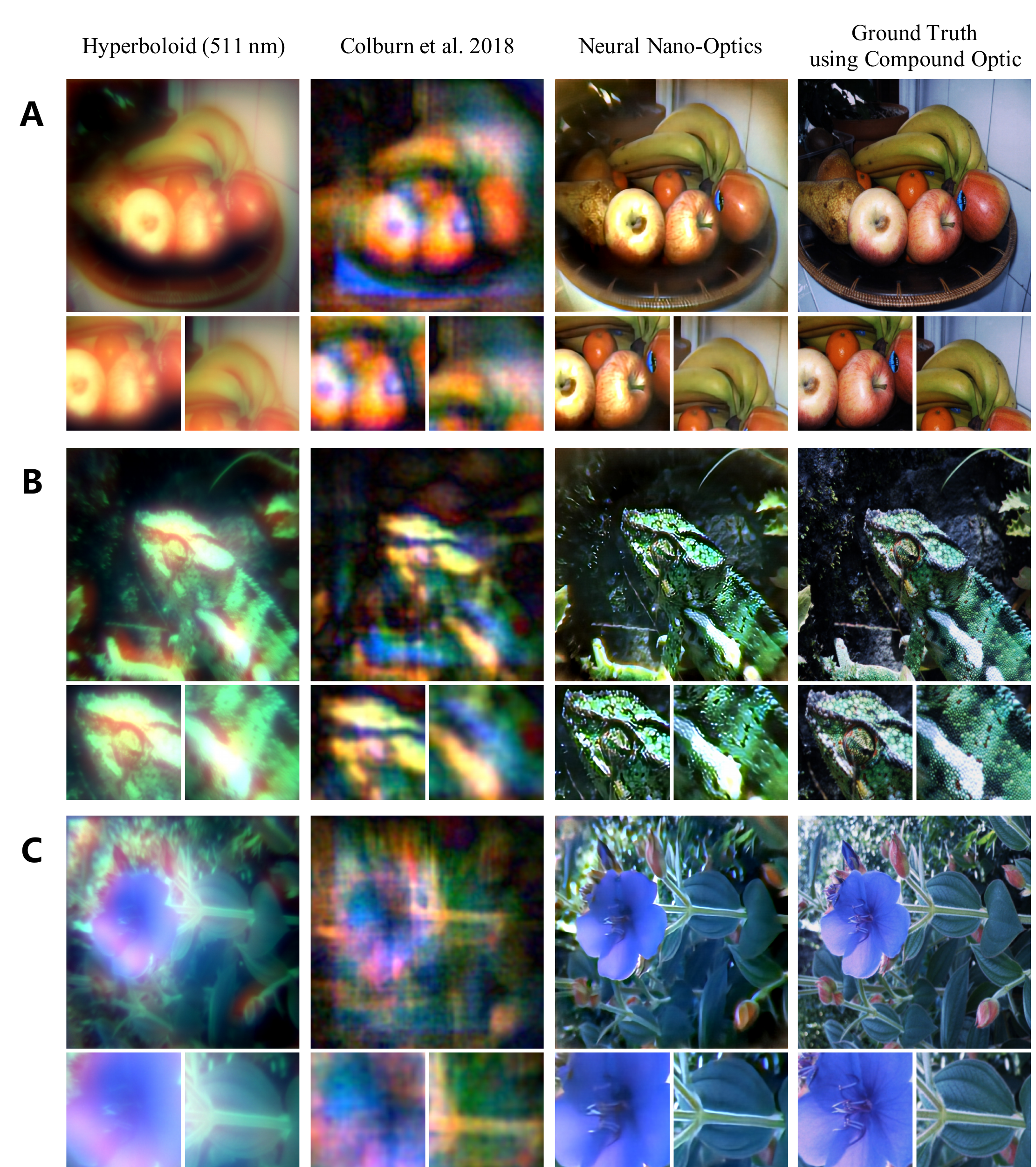}
    \caption{Experimental imaging results are shown in (A), (B), and (C), with insets included below each row. Compared to existing state-of-the-art designs, the proposed neural nano-optics produces high-quality wide FOV reconstructions corrected for aberrations. We compare our reconstructions to ground truth acquisitions using a high-quality, six-element compound refractive optic, and we demonstrate accurate reconstructions even though our meta-optic volume is \emph{550000$\times$ lower than the compound optic.}}
    \label{fig:results}
\end{figure*}

\begin{figure*}
    \centering
    \includegraphics[width=0.86\linewidth]{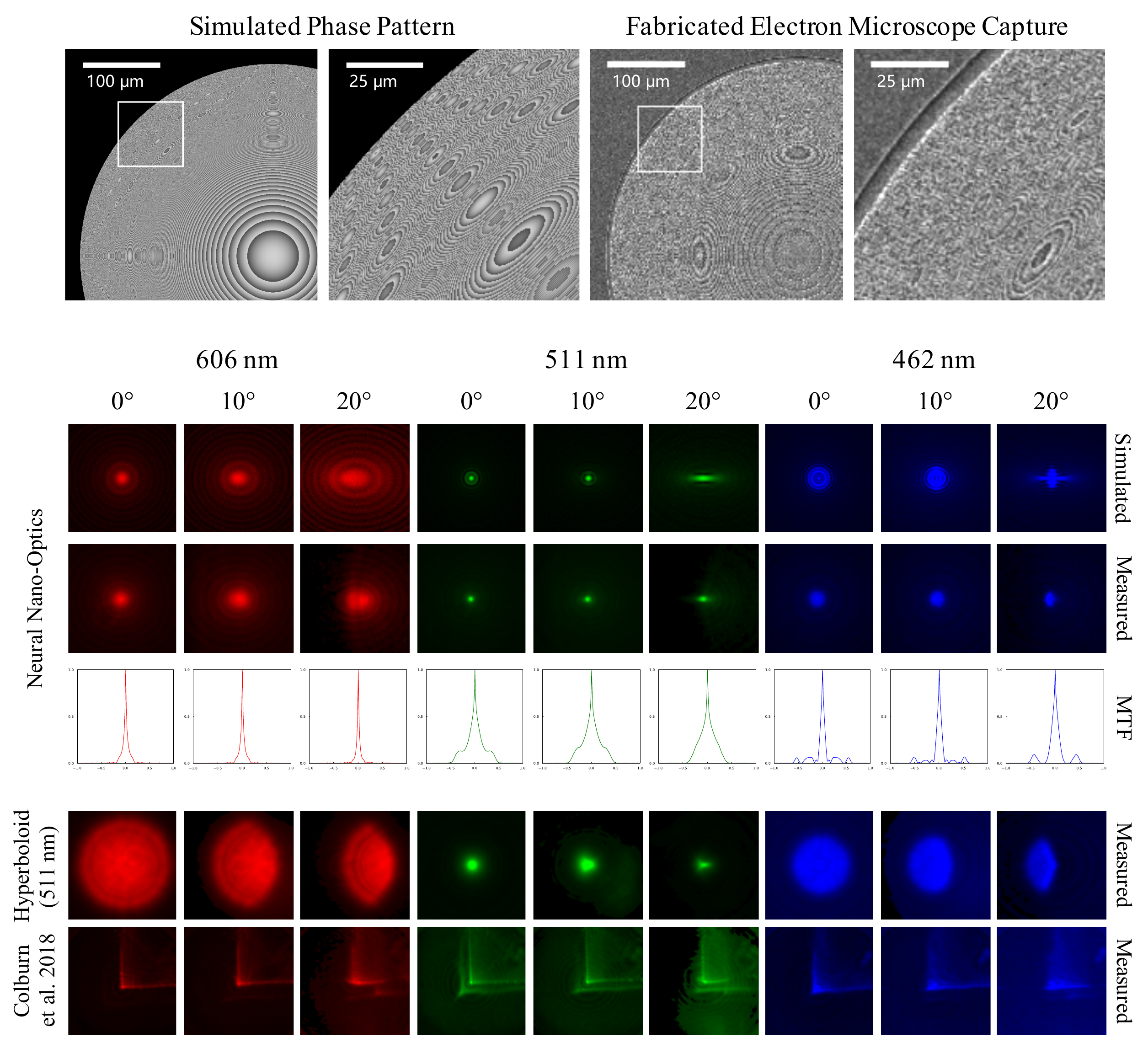}
    \caption{The proposed learned meta-optic is fabricated using electron-beam lithography and dry etching, and the corresponding measured PSFs, simulated PSFs, and simulated MTFs are shown. Before capturing images, we first use the fabricated optics to capture spatial PSFs to account for fabrication inaccuracies. Nevertheless, the match between the simulated PSFs and the captured PSFs demonstrates the accuracy of our metasurface proxy model. In contrast, the PSFs of the traditional meta-optic and the cubic design proposed by Colburn et al.\cite{Colburn2018MetasurfaceOF} demonstrate severe chromatic aberrations at the red and blue wavelengths and across the different field angles. The proposed learned design maintains consistent PSF shape across the visible spectrum and for all field angles across the FOV, facilitating downstream deconvolution and the final image reconstruction.}
    \label{fig:fab}
\end{figure*}